\documentclass[prd,preprintnumbers,amsmath,amssymb, nofootinbib]{revtex4}
\usepackage{amsmath,amssymb,graphicx}
\usepackage{epsf}
\usepackage{epstopdf}
\usepackage {amssymb}

\input{epsf.sty}

\newcommand{\head}[1]{\textnormal{\textbf{#1}}}
\usepackage{booktabs}
\def\beq{\begin{equation}}
\def\eeq{\end{equation}}
\def\bea{\begin{eqnarray}}
\def\eea{\end{eqnarray}}

 \def\no{\nonumber}

\begin{document}
\title{ \Large On the running of the spectral index to all orders: a new model dependent approach to constrain inflationary models}

\author{Moslem Zarei$^{2,3}$}
\email[]{m.zarei[AT]cc.iut.ac.ir}

\affiliation{$^{2}$Department of Physics, Isfahan University of
Technology, Isfahan 84156-83111, Iran}

\affiliation{$^{3}$ School of Astronomy, Institute for Research in Fundamental
Sciences (IPM), P. O. Box 19395-5531, Tehran, Iran}


\begin{abstract}

In conventional model independent approaches, the power spectrum of primordial perturbations is characterized by such free parameters as spectral index, its running, the running of running, and the tensor-to-scalar ratio. In this work we, show that, at least for simple inflationary potentials, one can find the primordial scalar and tensor power spectra exactly by resumming over all the running terms. In this model-dependent method, we expand the power spectra about the pivot scale to find the series terms as functions of the e-folding number for some single field models of inflation. Interestingly, for the viable models studied here, one can sum over all the terms and evaluate the exact form of the power spectra. This in turn gives more accurate parametrization of the specific models studied in this work. We finally compare our results with the recent cosmic microwave background data to find that our new power spectra are in good agreement with the data.

\end{abstract}


\maketitle

\section{ Introduction}

The recent precision data released by the Planck team favor inflation as a paradigm for explaining not only the temperature fluctuations in the Cosmic Microwave Background (CMB) but also large scale structure formation \cite{Ade:2015lrj,Ade:2015xua,Ade:2015tva}. As it is convenient, the Planck team considers a power-law form for the spectrum of primordial curvature perturbations as in \eqref{powerlaw} below,
\beq
\mathcal{P}_{\mathcal{R}}(k)=\mathcal{P}_{\mathcal{R}}(k_{\ast})\left(\frac{k}{k_{\ast}}\right)^{n_s-1}~,\label{powerlaw}
\eeq
with scalar amplitude, $\mathcal{P}_{\mathcal{R}}(k_{\ast})$, and spectral index, $n_s$, both defined at the pivot scale, $k_{\ast}$. Fitting with the recent Planck data \cite{Ade:2015lrj,Ade:2015xua} and the BICEP/Keck array \cite{Ade:2015tva,Array:2015xqh}  gives $n_s$, $\mathcal{P}_{\mathcal{R}}(k_{\ast})$, and tensor-to-scalar ratio $r$ at $k_{\ast} = 0.05 \textrm{Mpc}^{-1}$ as follows
\bea
&&n_s= 0.9655 ~ \pm ~ 0.0062 ~~~(68~ \%~ \textrm{CL}, \textrm{Planck TT+lowP}),~~~~~~~~~~~\ln\left(10^{10}\mathcal{P}_{\mathcal{R}}(k_{\ast})\right)=3.089 \pm 0.036~, \no \\&&
\textrm{and}\no \\&& r<0.07 \,(95\% \,\textrm{CL},\textrm{BICEP2/Keck})~. \label{r}
\eea

Considering $n_s$ and $r$ as model parameters, the Planck team first produced the marginalized joint distribution probability regions for these parameters. They then generated the parametric plot of the $n_s$ and $r$ parameters (predicted by different inflationary models) in the $n_s-r$ plane of Planck, whereby they applied the first method to distinguish between the models of inflation \cite{Ade:2015xua}. This analysis showed that CMB temperature fluctuations are consistent with the simplest inflationary models with plateau-like inflaton potentials with $V_{\phi\phi}<0$ \cite{Ade:2015xua}. These models are single fields and predict a nearly Gaussian scale invariant power spectrum. The Planck team also considered the running of the spectral index, $d\,n_s/d\ln k$, as the next correction to the power law power spectrum and found that, for $r=0$, the CMB data would favor the power spectrum with a negligible running \cite{Ade:2015xua}
\beq
d\,n_s  / d\ln k=-0.0084 \pm 0.0082 ~~(68~ \% ~\textrm{CL}, \textrm{Planck TT+lowP})~.
\eeq
This value just improves the best fit likelihood by $\Delta\chi^{2}\approx -0.8$ \cite{Ade:2015xua}. However, considering the running of running of the spectral index, $d^{2}\,n_s  / d\ln k^{2}$, the Planck TT+lowP data gives \cite{Ade:2015xua}
\bea
&&n_s=0.9569 \pm 0.0077~~(68~\%~\textrm{CL})~,\no \\ &&
d\,n_s  / d\ln k= 0.011^{+0.014}_{-0.013}~~(68~\%~\textrm{CL})~,\no \\ &&
d^{2}\,n_s  / d\ln k^{2}=  0.029^{+0.015}_{-0.016}~~(68~\%~\textrm{CL})~,
\eea
 which improves the best fit likelihood by $\Delta\chi^{2}\approx -4.8$ at low multipoles \cite{Ade:2015xua}.

Another approach explored by the Planck team to disfavor/favor models of inflation is based on the reconstruction of potential in the neighborhood of $\phi_{\ast}$, with the value of $\phi$ being determined at the moment when the mode $k_{\ast}$ exists the horizon for the first time \cite{Kinney:2005in,Kinney:2006qm,Peiris:2006ug,Lesgourgues:2007gp}. Using this method, one can constrain the observable part of the inflaton potential in the interval of $\phi$ corresponding to the scales of CMB. The observable parameters
$n_s$, $d\,n_s/d\ln k$, and $r$ are given as a function of the slow-roll parameters which are calculated at $k_{\ast}$. It will now be possible to include the slow roll parameters in the parameter estimation process. Using the slow roll flow equations \cite{Liddle:1994dx,Kinney:2002qn}, one can compute their values at all times or for any values of $k$. The slow roll hierarchy can be truncated at any given point and constraints can be imposed on the slow roll parameters while there are no direct constraints on the relevant parameters $n_s$, $r$, and $d\,n_s/d\ln k$. However, using the constraint on the slow roll parameters, one can constrain these parameters to every order in the slow roll. Moreover, the inflationary potential can be reconstructed to every order of Taylor series around $\phi_{\ast}$ by assuming a flat prior on the slow roll parameters \cite{Kinney:2005in,Kinney:2006qm,Peiris:2006ug,Lesgourgues:2007gp}.

The third approach proposed in \cite{Mortonson:2010er} considers the specific inflationary models with one or more free parameters. It is assumed that the reheating phase is also controlled by other free parameters. In this method, the power spectrum is calculated numerically without assuming slow roll approximation. This can be performed by adopting numerical codes for an efficient computation of inflationary perturbations. In order to detail the end of the inflation, one can consider at least three possibilities according to different entropy generation mechanisms \cite{Planck:2013jfk}. This method significantly reduces the theoretical uncertainty in the parameter space of inflationary models. However, the comparison between the constraints resulting from this method and those of the standard analysis with $n_s$, $r$, and $dn_s/d\ln k$ parameterizations is not straightforward.

In the present paper, a different approach in utilized to we constrain inflationary models although it has similarities with the method reported in \cite{Martin:2014vha} which is a model-dependent approach and in which the slow roll approximation was used to calculate the power spectrum for different inflationary models while the first few lowest-order terms are preserved. The slow roll parameters were also calculated in terms of the parameters of each model and the e-folding number $N$. The authors then compared the inflationary models with CMB data and scanned the parameter space of each model. This model dependent approach yielded more accurate parametrization for any model of inflation. In the present work, we also use the slow roll approximation and show that at least for some single field models, one can calculate the power spectrum in terms of $N$ by resumming over all the running terms.
 The main advantage of our work is that for these models we can calculate the scalar and tensor power spectra as a function of $k$ with the e-folding number, $N(k_{\ast})$, as the free parameter. In fact, we sum over the running terms to all orders. There are no spectral indices, running, the running of running, or tensor-to-scalar ratio parameters in the power spectra calculated in our method. Using the CMB data analysis, one can find the best fit value of e-folding number, $N(k_{\ast})$, for inflationary models and it will, therefore, be possible to favor the models of inflation. Additionally, we study the three types of single field inflation models. After fitting with the data, the constraint on $N(k_{\ast})$ is determined which will be used to calculate $n_s$ and the parameter $r$. It should noted that $n_s$ and $r$ derived in this method are all model-dependent while, in the conventional data fitting of the Planck team, those parameters characterizing the primordial power spectra are model-independent. Therefore, the meaning of the parameters derived by these two methods are certainly different.

The paper is organized as follows: In section II, we review the derivation of the scalar and tensor power spectra and find a representation for the spectral indices $n_s(k)$ and $n_t(k)$ by expanding about the pivot scale $k_{\ast}$. In section III, we discuss how the e-folding number, $N(k)$, is generally constrained in the inflationary paradigm. We then turn to the inflation models with chaotic, hill top, and Starobinsky potentials to find their power spectra as a function of $N(k_{\ast})$. In section IV, we present our CMB data analysis and compare the predictions of the new power spectra with the results obtained from the conventional $\Lambda$CDM model.

\section{The Running of spectral index }

We consider a single field inflation with the potential $V(\phi)$. By convention, the slow roll parameters are defined as follows \cite{Lyth:1998xn}
\beq
\epsilon_{V}=\frac{M^{2}_{Pl}}{2} \left(\frac{V'}{V}\right)^{2} \:\:\:\:\:\:\:\:\:\: \textrm{and} \:\:\:\:\:\:\:\:\:\: \eta_{V}=M^{2}_{Pl} \left(\frac{V''}{V}\right)~,\label{eps}
\eeq
where, the primes denote the derivative with respect to the inflaton field $\phi$. The slow roll condition is satisfied when $\epsilon_{V}\ll 1$ and $|\eta_{V}|\ll 1$.
The power spectrum of the curvature $\mathcal{R}$ can be Taylor expanded in terms of $\ln\! k$ around the pivot scale $k_{\ast}$ \cite{Lyth:1998xn}
\beq
\ln\mathcal{P}_{\mathcal{R}}(k)=\ln\mathcal{P}_{\mathcal{R}}(k_{\ast})+\frac{d\ln\mathcal{P}_{\mathcal{R}}(k_{\ast})}{d\ln k}\ln\frac{k}{k_{\ast}}+\frac{1}{2\,!} \frac{d^{2}\ln\mathcal{P}_{\mathcal{R}}(k_{\ast})}{d\ln k^{2}}\ln^{2}\frac{k}{k_{\ast}}+\frac{1}{3\,!}\frac{d^{3}\ln\mathcal{P}_{\mathcal{R}}(k_{\ast})}{d\ln k^{3}}\ln^{3}\frac{k}{k_{\ast}}+\cdot\cdot\cdot~.
\eeq
Using the second term of this expansion, the scalar spectral index is defined by $n_s(k_{\ast})\equiv 1+d\ln\mathcal{P}_{\mathcal{R}}(k_{\ast})/d\ln k $. Using the sequence of higher derivative terms, one can define the running and the running of the running of the spectral index as in \eqref{alpha} and \eqref{beta}, respectively
\bea
\alpha_s(k_{\ast})\equiv\frac{dn_s}{d\ln k}(k_{\ast})=\frac{d^{2}\ln\mathcal{P}_{\mathcal{R}}(k_{\ast})}{d\ln k^{2}}\label{alpha}~,\\
\beta_s(k_{\ast})\equiv\frac{d^{2}n_s}{d\ln k^{2}}(k_{\ast})=\frac{d^{3}\ln\mathcal{P}_{\mathcal{R}}(k_{\ast})}{d\ln k^{3}}~,\label{beta}
\eea
where, the running of running parameter, $\beta_s(k_{\ast})$, was first measured by Planck satellite \cite{Planck:2013jfk}. Using the same approach, one can expand the tensor power spectrum $\mathcal{P}_{t}(k)$ about the pivot scale $k_{\ast}$ as
\beq
\ln\mathcal{P}_{t}(k)=\ln\mathcal{P}_{t}(k_{\ast})+\frac{d\ln\mathcal{P}_{\mathcal{R}}(k_{\ast})}{d\ln k}\ln\frac{k}{k_{\ast}}+\frac{1}{2\,!} \frac{d^{2}\ln\mathcal{P}_{t}(k_{\ast})}{d\ln k^{2}}\ln^{2}\frac{k}{k_{\ast}}+\cdot\cdot\cdot~,
\eeq
where, $\mathcal{P}_{t}(k_{\ast})=r\,\mathcal{P}_{\mathcal{R}}(k_{\ast})$ and one can thus define the corresponding tensor spectral index $n_t(k_{\ast})\equiv d\ln\mathcal{P}_{t}(k_{\ast})/d\ln k$ and its running $\alpha_t(k_{\ast})\equiv d^{2}\ln\mathcal{P}_{t}(k_{\ast})/d\ln k^{2}$. On the other hand, the standard calculations in the slow roll regime give the curvature and tensor power spectra for single field models as follows
\bea
&&\mathcal{P}_{\mathcal{R}}(k)\simeq\frac{2}{3\pi M^{6}_{Pl}}\frac{V^{3}}{V'^2}~,\label{ps} \\&&
\mathcal{P}_{t}(k)\simeq\frac{16V}{3\pi M^{4}_{Pl}}~,\label{pt}
\eea
in which we have considered the power spectra up to the lowest powers of the slow roll parameters. Accurate computations lead to the power spectra including the next to the leading order corrections \cite{Stewart:1993bc}. Using the following relation
\beq \label{deriv}
\frac{d}{d\ln\,k}\simeq-M^{2}_{Pl}\,\frac{V'}{V}\,\frac{d}{d\phi}~,
\eeq
 it will be possible to derive the scalar and tensor spectral indices and their associated running terms in terms of the slow roll parameters up to the first order in the slow roll parameters. With straightforward calculations, one can show that
\bea
&&n_s(k_{\ast})\simeq1-6\epsilon_V+2\eta_V~,\\ &&
\alpha_s(k_{\ast})\simeq16\eta_{V}\epsilon_{V}-24\epsilon_{V}^2-2\xi^{2}_V~, \label{run1}\\&&
\beta_s(k_{\ast})\simeq192\epsilon^{3}_{V}-192\epsilon_{V}^2\eta_V+32\epsilon_V\eta^{2}_V+24\epsilon_V\xi^{2}_{V}-2\eta_V\xi^{2}_V-2\varpi_{V}^{3}~, \label{run2}
\eea
 where,
\beq
\xi^{2}_{V}=M^{4}_{Pl}\frac{V'\,V'''}{V^2}~,\:\:\:\:\:\:\:\:\:\: \textrm{and}  \:\:\:\:\:\:\:\:\:\:\varpi_{V}^{3}=M^{6}_{Pl}\frac{V'^2\,V''''}{V^3}~,\label{xi1}
\eeq
and for the tensor perturbations, we find \cite{Gorbunov:2011zzc}
\bea
&& n_t(k_{\ast})\simeq -2\epsilon_V~,\\&&
\alpha_t(k_{\ast})\simeq 4\eta_{V}\epsilon_{V}-8\epsilon_{V}^2 ~.
\eea
Now, the power spectrum of curvature and tensor perturbations can generally be deduced in the following forms
\bea \label{powers}
&&\ln\mathcal{P}_{\mathcal{R}}(k)=\ln\mathcal{P}_{\mathcal{R}}(k_{\ast})+(n_s(k_{\ast})-1)\ln\frac{k}{k_{\ast}}+\frac{1}{2\,!}\frac{dn_s(k_{\ast})}{d\ln k}\ln^{2}\frac{k}{k_{\ast}}+\frac{1}{3\,!}\frac{d^{2}n_s(k_{\ast})}{d\ln k^{2}}\ln^{3}\frac{k}{k_{\ast}}+\cdot\cdot\cdot~,
\\&&\ln\mathcal{P}_{t}(k)=\ln\mathcal{P}_{t}(k_{\ast})+n_t(k_{\ast})\ln\frac{k}{k_{\ast}}+\frac{1}{2\,!}\frac{dn_t(k_{\ast})}{d\ln k}\ln^{2}\frac{k}{k_{\ast}}+\cdot\cdot\cdot~,\label{powert}
\eea
where, the expansions \eqref{sumns} and \eqref{sumnt} are effectively induced for the scalar and tensor spectral indices respectively
\beq \label{sumns}
n_s(k)-1=n_s(k_{\ast})-1+\sum_{l=1}^{\infty}\frac{d^{l}n_s(k_{\ast})}{d\ln k^{l}}\frac{\ln^{l}\frac{k}{k_{\ast}}}{(l+1)!}~,
\eeq
and

\beq \label{sumnt}
n_t(k)=n_t(k_{\ast})+\sum_{l=1}^{\infty}\frac{d^{l}n_t(k_{\ast})}{d\ln k^{l}}\frac{\ln^{l}\frac{k}{k_{\ast}}}{(l+1)!}~.
\eeq
The parameters $n_s(k)$ and $n_t(k)$ are given based on the specific model of inflation. The series can be truncated by putting a cutoff on $l$. As an example, cutting off the series \eqref{sumns} at $l=2$  is identical to a scalar spectral index with the running and the running of running corrections. For a general $\Lambda$CDM model, the parameters such as $n_s(k_{\ast})$, $n_t(k_{\ast})$, $\alpha_s(k_{\ast})$ and $\beta_s(k_{\ast})$ are taken as free parameters with specific prior assumptions. However, as we will show, for some specific models of inflation, we can basically calculate these parameters in terms of the e-folding number $N_{\ast}\equiv N(k_{\ast})$. Hence, they are not all free parameters of the model. Given that in the $\Lambda$CDM model a set such as $(\mathcal{P}_{\mathcal{R}}(k_{\ast}), r, n_s, \alpha_s, \beta_s, \cdot\cdot\cdot)$ may represent the parameter space associated with the primordial perturbations, the second model-dependent approach gives the free parameters by the set $(\mathcal{P}_{\mathcal{R}}(k_{\ast}), N, \cdot\cdot\cdot)$ for some specific models of inflation. Here, the dots denote the basic set of other post-inflationary cosmological parameters.

 \section{  Calculation of the running of spectral index to all orders }

 As already described in the previous section, for a given inflationary model, the running of the spectral index can be calculated order by order. By the term, order by order, we mean that one can calculate the derivatives for each number of $l$ in the series \eqref{sumns} and \eqref{sumnt}. Accordingly, by summing over series terms, one can find the spectral indices with the running corrections to all orders. In order to find the series terms, we first have to write down the slow roll parameters and the spectral indices in terms of the e-folding number $N$. This parameter measures the accelerated expansion of universe during inflation from a given time $t$ where a mode passes the horizon to time $t_e$ denoting the end of inflation. For a single field model, $N$ is defined as \cite{Lyth:1998xn}
 \beq
 N(\phi)=\int_{t}^{t_e}H(t)dt=M^{-2}_{Pl}\int_{\phi_e}^{\phi}\frac{V}{V'}d\phi~,
 \eeq
where, $\phi_e$ is the physical clock that determines the end of inflation. Here, we consider three different kinds of single field inflationary potentials belong to the large field, small field, and the Starobinsky-like models. In the rest of this section, we derive the spectral indices for these models as a function of $N_{\ast}$. For a given model, we should first calculate the inflaton field $\phi$ as a function of $N$. It is plausible to find $\phi(N)$ for a wide range of inflation models \cite{Martin:2014vha}. The next step involves the calculation of $n_s$, $n_t$, and the running terms as a function of $N$ for each model. One property of the three models studied here is that one can compute the terms $d^{l}n_s(k_{\ast})/d\ln k^{l}$ and $d^{l}n_t(k_{\ast})/d\ln k^{l}$ in \eqref{sumns} and \eqref{sumnt} as functions of $n_s(k_{\ast})$, and $n_t(k_{\ast})$ respectively. This is the necessary condition for a given model to be studied by our method. One can factor out the spectral index terms to sum up the series \eqref{sumns} and \eqref{sumnt} in a simple way.

\subsection{ Chaotic potential }

To begin with, we consider the chaotic potential $V=g\phi^{n}$ from the large field model class where $n>0$ and $g$ is the coupling with the dimension $[g]=(\textrm{mass})^{4-n}$. As has been asserted, comparison with the Planck data disfavors the chaotic potential with $n=2$ although $n=1$ and $n=2/3$ models have better positions within the $95\% $ CL region \cite{Ade:2015xua}.
For the epoch $N\sim \mathcal{O}(10)$ before the end of inflation, $\phi\gg\phi_e$; one can, therefore, obtain the following result \cite{Peiris:2003ff,Gorbunov:2011zzc}
\beq
\phi(N)\simeq\sqrt{2nM^{2}_{Pl}N}~.\label{phiN}
\eeq
On the basis of this approximation and using \eqref{eps}, \eqref{ps}, \eqref{pt}, and \eqref{xi1}, we can calculate the scalar and tensor power spectra as well as the slow roll parameters as a function $N$
\bea
&&\mathcal{P}_{\mathcal{R}}(N)\simeq\frac{2g}{3\pi M^{6}_{Pl}n^{2}}\left(2nM^{2}_{Pl}N\right)^{\frac{n+2}{2}}~,\:\:\:\:\:\:\:\:\:
\mathcal{P}_{t}(N)\simeq\frac{16g}{3\pi M^{4}_{Pl}}\left(2nM^{2}_{Pl}N\right)^{\frac{n}{2}}~, \\&&
\epsilon_V\simeq\frac{n}{4N}~,\:\:\:\:\:\:\:\:\:\eta_V\simeq\frac{n-1}{2N}~,\:\:\:\:\:\:\:\:\:\xi_V^2\simeq\frac{(n-1)(n-2)}{4N^2}\:\:\:\:\:\textrm{and}\:\:\:\:\:
\varpi_V^3\simeq\frac{(n-1)(n-2)(n-3)}{8N^3}~.\label{epsilon}
\eea
We recall that the scalar and tensor spectral indices are given in terms of slow roll parameters or equally through the derivative of the power spectrum. Either way, one can show that
\beq
n_s-1\simeq-\frac{n+2}{2N_{\ast}}~,\:\:\:\:\:\:\:\:\:n_t\simeq -\frac{n}{2N_{\ast}} \:\:\:\:\:\:\:\textrm{and}\:\:\:\:\:\:\:r\simeq\frac{4n}{N_{\ast}}~,
\eeq
where, the tensor-to-scalar ratio is given by $r= 16\epsilon_V= -8n_t$. Using the same method, substitution of equation \eqref{epsilon} in \eqref{run1} and \eqref{run2} gives $\alpha_{s}$ and $\beta_{s}$ as in the following
\bea
\alpha_{s}\simeq\frac{1!\,(n_s-1)}{N_{\ast}} ~,\:\:\:\:\:\:\:\:\:
\beta_{s}\simeq\frac{2!\,(n_s-1)}{N_{\ast}^{2}}~.
\eea
Using equations \eqref{deriv} and \eqref{phiN} as well as the derivatives of the power spectrum \eqref{alpha} and \eqref{beta}, we obtain the same results for $\alpha_{s}$ and $\beta_{s}$. We can continue to derive the higher derivatives of the spectral index using this approach. After some computations, the following general result is given \cite{Kosowsky:1995aa}
\bea \label{dns}
\frac{d^{l}n_s}{d\ln k^{\textit{l}}}\simeq \frac{l\,!\,(n_s-1)}{N_{\ast}^{l}}~.
\eea
In a similar manner, for the tensor spectral index, we have
\beq \label{dnt}
\frac{d^{l}n_t}{d\ln k^{l}}\simeq\frac{l!\,n_t}{N_{\ast}^{l}}~.
\eeq
Returning to the series \eqref{sumns} and \eqref{sumnt}, we see that substituting the results \eqref{dns} and \eqref{dnt} in the relevant equations and making the summation will yield
\bea
n_s(k)-1&\simeq& (n_s(k_{\ast})-1)\sum_{l=0}^{\infty}\frac{x^{l}}{l+1}\no \\ && \!\!\!\!\!\!\!\!=\frac{n+2}{2N_{\ast}}\frac{\ln(1-x)}{x}~, \label{nschaotic}
\eea
and
\bea
n_t(k)&\simeq&\frac{n}{2N_{\ast}}\frac{\ln(1-x)}{x}~,\label{ntchaotic}
\eea
where, $x\equiv \ln(k/k_{\ast})/N_{\ast}$. The tensor-to-scalar ratio can in general be determined as a function of $k$; however, we define it at the point of $k=k_{\ast}$ with no running
\bea
r&=&-8\,n_t(k_{\ast})\simeq  \frac{4\,n}{N_{\ast}}~.
\eea
The key ingredient in deriving the parameters $n_s(k)$, $n_t(k)$, and $r$ is, therefore, the model-dependent parameter $N_{\ast}$ which can practically be regarded as a free parameter of the model. Compared with the conventional model-independent approach, there are no spectral indices, running corrections, or  tensor-to-scalar ratio in this case but only the e-folding number $N_{\ast}$ takes part in the parameter space. This point will be further investigated by data analysis at the end of this paper.

\subsection{ Starobinsky Model  }

Another class of inflationary models involves the extension of theories of gravity up to the $R^2$ term known as $R^2$ inflation \cite{Starobinsky:1980te,Mukhanov:1981xt}
It has been shown that, in the Jordan frame this model can be described by the following potential \cite{Maeda:1987xf}
\beq
V(\phi)=\frac{1}{8\,\alpha}\left(1-e^{-\sqrt{2/3}\,\phi}\right)^{2}~,
\eeq
where, $\alpha$ is the coupling of $R^2$ term. There is an excellent agreement between this model and the Planck data \cite{Ade:2015xua}. This is due to the small value of $r$ predicted by model, which indicates the tiny gravitational waves. For this potential, one can also show that \cite{Lyth:2009zz,Hinshaw:2012aka}
\beq
\phi(N)\simeq \sqrt{\frac{3}{2}}\ln\frac{4N}{3}~,\:\:\:\:\:\:\:\:\:\:\:\:\:\:n_s-1\simeq-\frac{2}{N_{\ast}}\:\:\:\:\:\:\: \textrm{and} \:\:\:\:\:\:\: n_t\simeq -\frac{3}{2N_{\ast}^2}~.
\eeq
Employing the method described above yields
\beq
\frac{dn_s}{d\ln k}\simeq -\frac{2}{N_{\ast}^{2}}~,\:\:\:\:\:\:\:\:\:\frac{d^{2}n_s}{d\ln k^{2}}\simeq -\frac{4}{\,N_{\ast}^{3}}\:\:\:\:\:\:\:\textrm{and}\:\:\:\:\:\:\:
\frac{d^{l}n_s}{d\ln k^{l}}\simeq\frac{l!\,(n_s-1)}{\,N_{\ast}^{l}}~.
\eeq
Furthermore, we can calculate
\beq
\frac{dn_t}{d\ln k}\simeq -\frac{3}{N_{\ast}^{3}}~,\:\:\:\:\:\:\:\:\:\frac{d^{2}n_t}{d\ln k^{2}}\simeq -\frac{9}{\,N_{\ast}^{4}}\:\:\:\:\:\:\:\textrm{and}\:\:\:\:\:\:\:
\frac{d^{l}n_t}{d\ln k^{l}}\simeq \frac{(l+1)!\,(n_t)}{\,N_{\ast}^{l}}~.
\eeq
Therefore, using the series \eqref{sumns} and \eqref{sumnt}, we find
\bea
n_s(k)-1\simeq
\frac{2}{N_{\ast}}\frac{\ln(1-x)}{x}, \:\:\:\:\:\:\:\:\:\:\:\:\:\:\textrm{ and}  \:\:\:\:\:\:\:\:\:\:\:\:\:\: n_t(k)\simeq-\frac{3}{2N^{2}_{\ast}}\frac{1}{(1-x)}~,\label{nsstarobinsky}
\eea
and the tensor-to-scalar ratio for this case will be $r\approx12/N^{2}_{\ast}$. Here, the $k$-dependence of the scalar spectral index is the same as that if  the chaotic potential except that new forms are derived for the $n_t(k)$ and $r$. Similar to the previous example, the parameter $N_{\ast}$ remains as a key element in the parameter space.

\subsection{ Hill top potential }

In what follows, an example will be examined from the small field models and the symmetry breaking hill top inflaton potential will be investigated
\beq
V=\Lambda^{4}\left(1-\frac{\phi^{p}}{\mu^{p}}\right)~,
\eeq
where, $\Lambda$ is the coupling constant and $\mu$ is the value of inflaton at the minimum of the potential. We are interested in the potentials with $p\geq 3$. In this model, the slow roll parameters and scalar spectral tilt are given in terms of the number of inflationary e-folds as follows \cite{Peiris:2003ff}
\beq
\phi(N) \simeq \left(\frac{\mu ^p}{p (p-2)M^2_{Pl} N }\right)^{\frac{1}{p-2}}~,\:\:\:\:\:\:\:\:\:\epsilon_V\simeq\frac{p^2}{2}\frac{M^{2}_{Pl}}{\phi^{2}}\left(\frac{\phi}{\mu}\right)^{2p}~,\:\:\:\:\:\:\:\:\: \eta_V\simeq-p(p-1)\frac{M^{2}_{Pl}}{\phi^{2}}\left(\frac{\phi}{\mu}\right)^{p}~,\:\:\:\:\:\:\:\:\:n_s-1\simeq-\frac{p-1}{p-2}\frac{2}{N_{\ast}}~.
\eeq
Moreover, one can calculate
\beq
\frac{dn_s}{d\ln k}\simeq-\frac{p-1}{p-2}\frac{2}{N_{\ast}^{2}}~,\:\:\:\:\:\:\:\:\:
\frac{d^{2}n_s}{d\ln k^{2}}\simeq-\frac{p-1}{p-2}\frac{4}{N_{\ast}^{3}}~,\:\:\:\:\:\:\:\:\:\:\textrm{and}\:\:\:\:\:\:\:\:\:\:\frac{d^{l}n_s}{d\ln k^{l}}\simeq\frac{l!\,(n_s(k_{\ast})-1)}{N_{\ast}^{l}(k_{\ast})}~.
\eeq
Therefore, the running of scalar spectral index to all orders is given by
\bea
n_s(k)-1&\simeq&\frac{p-1}{p-2}\frac{2}{N_{\ast}}\frac{\ln(1-x)}{x}~. \label{nshilltop}
\eea
The spectral index of tensor perturbations is given in as the following
\bea
n_t&\simeq&-\frac{p^2}{\hat{\mu}^{2}}\left[\frac{\hat{\mu}^{2}}{p(p-2)N_{\ast}}\right]^{\frac{2(p-1)}{p-2}}~.
\eea
where, $\hat{\mu}=\mu/M_{Pl}$. Hence, the running of $n_t$ to all orders is given by
\bea
n_t(k)&\simeq & n_t(k_{\ast})\left[1+\frac{2(p-1)}{2!\,(p-2)}\,x+\frac{2(p-1)(3p-4)}{3!\,(p-2)^{2}}\,x^{2} +\frac{4(p-1)(2p-3)(3p-4)}{4!\,(p-2)^{3}}\,x^{3}
\right. \nonumber \\ &&\left. \:\:\:\:\:\:\:\:\:\:\:\:\:\:\:\:\:\:+\frac{4(p-1) (2 p-3) (3 p-4) ( 5 p-8)}{5!\,(p-2)^{4}}\,x^{4} +\cdot\cdot\cdot\right] \nonumber \\ &&\!\!\!\!\!\!\!=n_t(k_{\ast})\frac{p-2}{p\,x}\left[1-(1-x)^{\frac{-p}{p-2}}\right]~.\label{nthilltop}
\eea
The tensor to scalar ratio parameter is fixed using the relation $r=-8n_t(k_{\ast})$. Interestingly, the $n_s(k)$ parameter predicted by the hill-top potential is the same as those given by the Starobinsky and chaotic potentials. However, $n_t$ and $r$ in this model have new forms and contain the dimensionless parameter $\hat{\mu}$ in addition to the $N_{\ast}$. Therefore, the hill-top model contributes one more free parameter to the parameter space.

\section{The CMB data analysis }

So far we have computed $n_s(k)$, $n_t(k)$, and $r$ as a function of number of e-folding $N_{\ast}$ and the parameters of inflaton potentials.
 Despite the possible uncertainties, one can impose upper and lower bounds on $N_{\ast}$. The main uncertainty is due to the stage of reheating. However, one can assume an instantaneous reheating to gain a maximum value for $N_{\ast}$. For a comoving scale $k$ which crosses the Hubble radius during inflation, the e-folding number $N(k)$ is given by \cite{Liddle:2003as,Dodelson:2003vq}
\beq
N(k)= -\ln\frac{k}{a_0H_0}+\frac{1}{3(1+w_{\textrm{reh}})}\ln\frac{\rho_{\textrm{reh}}}{\rho_{\textrm{end}}}+
\frac{1}{4}\ln\frac{\rho_{\textrm{eq}}}{\rho_{\textrm{reh}}}
+\frac{1}{2}\ln\frac{ M_{Pl}^{4}}{2\,\rho_{\textrm{eq}}}+\frac{1}{2}\ln 3\pi^{2}r\mathcal{P}_{\mathcal{R}}(k_{\ast})+\ln 219\Omega_{0}h
\eeq
where, $a_0H_0$ is the present horizon scale, $\rho_{\textrm{end}}$ is the energy density at the end of inflation, $\rho_{\textrm{reh}}$ is the energy density of universe at the reheating stage, $H_{\textrm{eq}}$ denotes the energy scale of equality, and $w_{\textrm{reh}}$ is the effective equation of state between the energy scales $\rho_{\textrm{end}}$ and $\rho_{\textrm{reh}}$. For an extreme case where reheating is instantaneous, $\rho_{\textrm{end}}=\rho_{\textrm{reh}}$, and where there is no significant reduction in energy density of inflation, so that $\rho_{\textrm{end}}=3\pi^{2}M_{Pl}^{4}r\mathcal{P}_{\mathcal{R}}(k_{\ast})/2$, we find a maximum e-folding number as follows
\beq
N(k)= 68.3-\ln\frac{k}{a_0H_0}+\frac{1}{4}\ln r \mathcal{P}_{\mathcal{R}}(k_{\ast})~,
\eeq
which gives $N(k_0)\simeq 63$ for our present Hubble scale $k_{0}=a_0H_0$ and with $r=0.2$ and $\mathcal{P}_{\mathcal{R}}(k_{\ast})=2.2\times 10^{-9}$. Nevertheless, for a prolonged reheating process, the expected number of reheating will be reduced. Assuming $0\leq w_{\textrm{reh}}< 1/3$ for the reheating phase, we can write
\beq
N(k_0)=63+\frac{1-3w_{\textrm{reh}}}{12(1+w_{\textrm{reh}})}\ln\frac{\rho_{\textrm{reh}}}{\rho_{\textrm{end}}}~.
\eeq
The reheating energy density is defined in terms of reheating temperature as $\rho_{\textrm{reh}}=\pi^{2}g_{\textrm{reh}}T_{\textrm{reh}}/30$ where $g_{\textrm{reh}}$, the effective number of relativistic species at the end of inflation, is usually set to $g_{\textrm{reh}}\simeq 100$. We now rewrite $N(k_0)$ in terms of reheating temperature
\beq
N(k_0)=63+\frac{1-3w_{\textrm{reh}}}{3(1+w_{\textrm{reh}})}\ln\frac{T_{\textrm{reh}}}{10^{16.5}\,\textrm{GeV}}~, \label{efold1}
\eeq
where, for an instantaneous reheating and with $r=0.2$, we obtain the previous result again. If the reheating process continues to the electroweak era such that the reheating temperature drops to $T_{\textrm{reh}}\sim 10^2\,\textrm{GeV}$, the reduction in the e-folding number will be
\beq
N(k_0)\simeq 63-\frac{11(1-3w_{\textrm{reh}})}{1+w_{\textrm{reh}}}~,
\eeq
and for $w_{\textrm{reh}}=0$, we find the minimum value of $N(k_0)\simeq 52$. In the same way, if reheating lasts to the nucleosynthesis epoch with $T_{\textrm{reh}}\sim 10^{-3}\,\textrm{GeV}$, we find $N(k_0)\simeq 48$ by assuming $w_{\textrm{reh}}=0$. On the other hand, as we can see from (\ref{efold1}), the equation of state $w_{\textrm{reh}}>1/3$ makes the e-folding number larger. However, this possibility requires somewhat exotic model building. As a result, for our present Hubble scale, $k_0$, the e-folding number is bounded between $48-63$. For the smaller scales such as the pivot scale $k_{\ast}=0.05$ which has always been of interest in the literature, $N_{\ast}$ is confined between $44-59$. This knowledge about the limits on $N_{\ast}$ is the key ingredient for our next data analysis. In the following, we compute the Bayesian probability distribution of models in order to obtain the constraints on the parameters of the models and $N_{\ast}$ by maximizing the likelihood functions. The bound on $N_{\ast}$ should be consistent with the reheating constraint.

After computing the $n_s(k)$ and $n_t(k)$ followed by $r$ for a given model, it will be possible to parameterize the primordial power spectra of the curvature and tensor perturbations as in the following
\beq
\mathcal{P}_{\mathcal{R}}(k)=\mathcal{P}_{\mathcal{R}}(k_{\ast})\left(\frac{k}{k_{\ast}}\right)^{n_s(k)-1}\:\:\:\:\:\:\:\:\:\:\textrm{and}\:\:\:\:\:\:\:\:\:\:
\mathcal{P}_{t}(k)=r\mathcal{P}_{\mathcal{R}}(k_{\ast})\left(\frac{k}{k_{\ast}}\right)^{n_t(k)}~, \label{powerst}
\eeq
 where, as we already mentioned, the key parameter is the e-folding number $N_{\ast}$. Therefore, unlike in the conventional method, fitting relation \eqref{powerst} with the data in this model does not directly constrain the pair $(n_s-r)$. Using the statistical Markov Chain Monte Carlo (MCMC) technique, it will be possible to evaluate the likelihood function, $\mathcal{L}$, and estimate the best fit of the cosmological parameters using CMB data. We adopt the modified version of the CAMB \cite{Lewis:1999bs} and CosmoMC \cite{Lewis:2002ah} codes to find the regions of the parameter space with the best fit. The CAMB code generates the CMB angular power spectra from the primordial curvature and tensor power spectrum. The modification of CAMB code is implemented in order to include the power spectra derived throughout this paper. One can then discriminate between various inflationary models discussed above by analyzing the data and finding the minimum effective $\chi^{2}\equiv -2\ln\mathcal{L}_{\textrm{max}}$ for each model. We investigate the constraints on model parameters, in particular $N_{\ast}$. By fixing $N_{\ast}$, it will be possible to calculate the spectral index $n_s$ or $r$ as a function of $N_{\ast}$. This only meant for comparison with the results of $\Lambda$CDM model obtained directly from an MCMC analysis where the $n_s-r$ couple is varied in the chain. We perform the numerical analysis for each inflationary model described above by comparing their predictions with the Planck TT + lowP and also the BICEP2, the Keck Array, and the Planck (BKP) data combinations. So far, we have calculated the spectral indexes $n_s(k)$ and $n_t(k)$ for three different classes of inflationary models. In order to perform the MCMC fit to the data, we first change the CAMB to include the new power spectra with $n_s(k)$, $n_t(k)$ as given in equations \eqref{nschaotic}, \eqref{ntchaotic}, \eqref{nsstarobinsky}, \eqref{nshilltop}, and \eqref{nthilltop}. It should also be noted that for each model, we have to take into account its relevant $r$ parameter computed above.

The $\Lambda$CDM model is explored by the parameter set $(\Omega_{b}h^{2}, \Omega_{c}h^{2}, \tau, 100\theta)$ where $\Omega_{b}h^{2}$, $\Omega_{c}h^{2}$ are the baryon and cold dark matter densities today, respectively, $\tau$ and $100\theta$ are the optical depth and the angular size of the sound horizon at the time of last-scattering, respectively. The primordial perturbations are parameterized by the set $(\mathcal{P}_{\mathcal{R}}(k_{\ast}),\: N_{\ast})$ in the chaotic and starobinsky models, and the set $(\mathcal{P}_{\mathcal{R}}(k_{\ast}),\: N_{\ast},\: \hat{\mu})$ in the hill top model. In our chains, we consider the flat priors just on the primordial perturbation parameters. We fix the post-inflationary parameters to the typical values consistent with the Planck 2015 data, namely $\Omega_{b}h^{2}=0.022$, $\Omega_{c}h^{2}=0.12$, $\tau=0.1$ and $100\theta=1.041$ \cite{Ade:2015lrj}. The primordial Helium fraction is also fixed to $Y_{P}=0.24$ \cite{Ade:2015lrj}. We impose a flatness condition $\Omega_{k}=0$, an adiabatic initial condition for inflation, and a massive neutrino with $N_{\nu}=3.046$. The MCMC chains are also run with the Gelman and Rubin criterion $R-1=0.02$ \cite{Rubin}. We have also fixed the pivot scale to $k_{\ast}=0.05\: \textrm{Mpc}^{-1}$.

 \begin{table}[h]
\centering 
\begin{tabular}{cc|ccccccc} \toprule[.5pt]
  \head{Models}  &&
\begin{tabular}{c} $-2\ln\,\mathcal{L}_{\textrm{max}}$  \\ Planck+lowP \:\:\:BKP \\
\end{tabular} &&
\begin{tabular}{c}  $N(k_{\ast})$  \\Planck+lowP \:\:\:BKP \\
\end{tabular} && \begin{tabular}{c}  $n_s(k_{\ast})$  \\ Planck+lowP \:\:\: BKP  \\
\end{tabular} && \begin{tabular}{c}  $r(k_{\ast})$  \\Planck+lowP \:\:\:BKP  \\
\end{tabular}   \\ \hline
  \midrule
\textrm{Chaotic} &
\begin{tabular}{c}$n=$2/3 \\ $n=$1 \\ $n=$2  \\
\end{tabular} &
\begin{tabular}{c}5635\:\:\:\:\:\:\:\:\:\:\:\:\:5657 \\ 5636\:\:\:\:\:\:\:\:\:\:\:\:\:5659 \\ 5637\:\:\:\:\:\:\:\:\:\:\:\:\:5658 \\
\end{tabular}&& \begin{tabular}{c}$32^{+12}_{-8}$\:\:\:\:\:\:\:\:\:\:\:$31^{+15}_{-7}$  \\ $32^{+15}_{-8}$\:\:\:\:\:\:\:\:\:\:\:$30^{+16}_{-8}$ \\ $66^{+27}_{-16}$\:\:\:\:\:\:\:\:\:\:\:$34^{+11}_{-10}$  \\
\end{tabular}&& \begin{tabular}{c}$0.96\pm0.01$\:\:\:\:\:\:\:\:\:\:\:$0.96\pm0.01$ \\ $0.95^{+0.02}_{-0.01}$\:\:\:\:\:\:\:\:\:\:\:$0.95^{+0.02}_{-0.01}$ \\ $0.97\pm 0.01$\:\:\:\:\:\:\:\:\:\:\:$0.94^{+0.02}_{-0.02}$  \\
\end{tabular}&& \begin{tabular}{c} $<0.1$\:\:\:\:\:\:\:\:\:\:\:\:\:\:\:\:$0.1$ \\ $<0.1$\:\:\:\:\:\:\:\:\:\:\:\:\:\:\:\:$0.1$ \\ $<0.1$\:\:\:\:\:\:\:\:\:\:\:\:\:\:\:\:$0.2$ \\
\end{tabular}
 \\ \hline
\textrm{Hilltop} &
\begin{tabular}{c} p=3 \\  p=4  \\
\end{tabular} &
\begin{tabular}{c} 5633\tiny{($\hat{\mu}$=9.5)} \:\:\:\:\:\:\:\:\:\:\:\normalsize 5656\tiny{($\hat{\mu}$=15)}  \\  5633\tiny{($\hat{\mu}$=12)}\:\:\:\:\:\:\:\:\:\:\:\:\normalsize 5657\tiny{($\hat{\mu}$=4.6)} \\
\end{tabular}&&
\begin{tabular}{c} $115^{+30}_{-26}$\:\:\:\:\:\:\:\: $89^{+42}_{-22}$ \\  $85^{+50}_{-19}$\:\:\:\:\:\:\:\:\:\: $91^{+45}_{-24}$ \\
\end{tabular}&&
\begin{tabular}{c} $0.95^{+0.01}_{-0.02}$\:\:\:\:\:\:\:\:\:\: $0.96^{+0.01}_{-0.02}$ \\  $0.98\pm 0.01$\:\:\:\:\:\:\:\:\:\: $0.98\pm 0.01$ \\
\end{tabular} &&
\begin{tabular}{c} $<0.004$\:\:\:\:\:\:\:\:\:\:\: $< 0.001$ \\  $<0.01$\:\:\:\:\:\:\:\:$<0.0002$\\
\end{tabular}  \\ \hline
\textrm{Starobisky} &
 &5633\:\:\:\:\:\:\:\:\:\:\:\:\:5657
&& $63^{+19}_{-18}$\:\:\:\:\:\:\:\:\:\:\:\: $58^{+24}_{-14}$
&& $0.97\pm 0.01$\:\:\:\:\:\:\:\:\:\: $0.97\pm 0.01$
 && $<0.003$\:\:\:\:\:\:\:\:$<0.004$
    \\ \hline\bottomrule[.5pt]
\end{tabular}
\label{tab:PPer}
\caption{\footnotesize   Marginalized $1\sigma$ confidence level limits for the parameters of inflationary models in Planck TT + lowP and BKP data set combinations, given with respect to the mean value. Note that the best fit value of $\mathcal{P}_{\mathcal{R}}(k_{\ast})=2.2\times 10^{-9}$ is found for all the models investigated. }
\end{table}

\begin{table}[h]
\centering 
\begin{tabular}{cc|ccccccccc} \toprule[.5pt]
  \head{Data}  &&
\begin{tabular}{c} $-2\ln\,\mathcal{L}_{\textrm{max}}$
\end{tabular} &&
\begin{tabular}{c}  $n_{s}(k_{\ast})$
\end{tabular} && \begin{tabular}{c}  $\alpha_s(k_{\ast})$
\end{tabular}&& \begin{tabular}{c}  $\beta_s(k_{\ast})$
\end{tabular} && \begin{tabular}{c}  $r(k_{\ast})$
\end{tabular}   \\ \hline
  \textrm{Planck+lowP } &
 &5634
&& $0.96\pm 0.01$
&& $\approx 2\times 10^{-2}$
&& $\approx 3\times 10^{-2}$
 && $<0.01$
  \\ \hline
BKP &
 &5657
&& $0.96\pm 0.01$
&& $\approx 1\times 10^{-2}$
&& $\approx 4\times 10^{-2}$
 && $<0.03$
  \\ \hline\bottomrule[.5pt]
\end{tabular}
\label{tab:PPer}
\caption{\footnotesize Marginalized $1\sigma$ confidence level limits for the parameters of the $\Lambda$CDM model in Planck TT + lowP and BKP
data set combinations, with respect to the mean value. For this model, we also find the best fit value of $\mathcal{P}_{\mathcal{R}}(k_{\ast})=2.2\times 10^{-9}$.}
\end{table}

The results of our analysis are summarized in Table I where the $\chi^{2}$ and constraint on $N_{\ast}$ are shown for different inflationary models. Using the value of $N_{\ast}$ given for each model, we have computed the $n_s$ and $r$ parameters for comparison with the results of the $\Lambda$CDM model summarized in Table II. It should be noted that $n_s$ and $r$ in Tables I and II certainly mean different things. In the following, $\Lambda$CDM model represents a cosmology with the primordial curvature perturbation parameters $(\mathcal{P}_{\mathcal{R}}(k_{\ast}), r, n_{s}(k_{\ast}), \alpha_s(k_{\ast}), \beta_s(k_{\ast}))$ and with the post-inflationary parameters as quantified above. In our data analysis, the best fit value of the amplitude $\mathcal{P}_{\mathcal{R}}(k_{\ast})$ is found to be $\mathcal{P}_{\mathcal{R}}(k_{\ast})=2.2\times 10^{-9}$ for all the inflationary models as well as for the $\Lambda$CDM one. It should be noted that the results in Table II for $r$ are new in the literature since the Planck 2015 analysis of $r$ only takes into account, at most, $\alpha_s$ but does not $\beta_s$.

 The chaotic potential was studied for $n=1,\:2/3,\:2$. The powers $n=1,\: 2/3$ are interesting in monodromy inflation models \cite{Flauger:2009ab,Silverstein:2008sg}. The constraint on the number of e-folding $N_{\ast}$ is obtained using the Planck TT + lowP and the BKP data sets. We find large error bars on $N_{\ast}$ because we have summed over all the running terms while the current data cannot constrain the running terms beyond $\beta_{s}$. The best fit point of $N_{\ast}$ for the $n=2$ model lies outside the expected range of $44-59$ for the $k_{\ast}$ scale. Moreover, taking into account the goodness of fit parameter, $\chi^{2}$, the $n=1,\:2/3$ model gives better fits to the Planck + lowP data. The $n=4$ inflation model is disfavored due to its worse $\chi^{2}$ value and prediction of large value for $N_{\ast}$; hence is not reported here. The spectral indices $n_{s}(k_{\ast})$ and $r(k_{\ast})$ can be calculated using the best fit values of $N_{\ast}$. It should be noted again that the these parameters are different in nature from those considered by the $\Lambda$CDM model. We can go one step further and compute the $\alpha_s(k_{\ast})$ and $\beta_s(k_{\ast})$ parameters for each model. Using this procedure, we are now able to compare among these models of inflation and the best fits of the $\Lambda$CDM model provided in Table II.
The results of the hill top potential have also been reported concisely in Table I. The parameter space for this model is extended by one more parameter $\hat{\mu}$, which justified the reduced value of $\chi^{2}$ for this model. However, the best fit values of $N_{\ast}$ are not within the expected range $44<N_{\ast}<59$ for either of the $p=3\: \textrm{or}\: 4$ models.
For the Starobinsky model, we have a parameter space as large as that of the chaotic model. However, we find a better fit with the Planck TT + lowP data. This agreement is due to the the small gravity wave predicted by this model. When we use the combined BKP data, $\chi^{2}$ becomes comparable with the corresponding parameter in the other models. It is known that the reheating temperatures allowed by the Starobinsky model is $T_{\textrm{reh}}\sim 3\times 10^{19}\: \textrm{GeV}$ \cite{Gorbunov:2010bn}, which gives $N_{\ast}\sim 54.4$. Considering the $1\:\sigma$ uncertainty, this theoretical prediction is satisfied by the results reported in Table 1 for the Starobinsky model.

\begin{figure}
\begin{center}\vspace{.5in}
\includegraphics[height=3in,width=5in,angle=0]{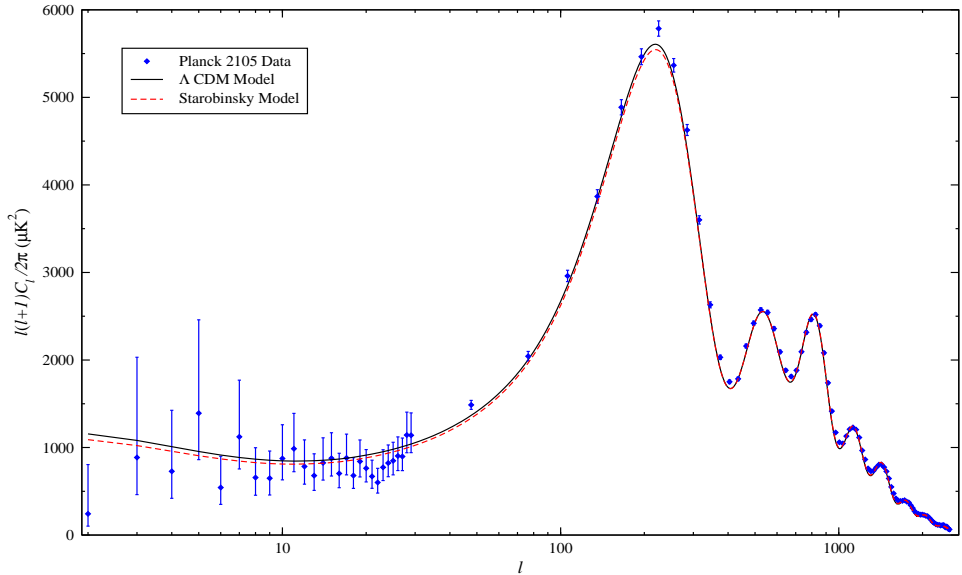}
\caption{\small \sl The angular TT power spectrum from the best fit of $\Lambda$CDM model (solid black curve) in comparison with the Starobinsky potential (dashed red curve) for the parameter values quoted in Tables I and II given using the Planck TT + lowP data. The blue diamonds show the Planck 2015 data.   \label{fig:Stupendous}}
\end{center}
\end{figure}

For the $\Lambda$CDM model, we estimate the scalar and tensor power spectra using equations \eqref{powers} and \eqref{powert}. With the Planck TT + lowP and BKP data sets, one can constrain the spectral index $n_{s}(k_{\ast})$, the running of the spectral index $\alpha_s(k_{\ast})$, and the running of running of the spectral index $\beta_s(k_{\ast})$. The results are summarized in Table II. Clearly, $\chi^{2}$ is better in this case than the corresponding values for the chaotic potential presented in Table I. This can be explained by the presence of more free parameters. For instance, in comparison with the chaotic potential, the $\Lambda$CDM model has 3 additional parameters. Interestingly, the Starobinsky model shows a better fit with fewer free parameters. Using two data sets, we have also found negligible values for running, $\alpha_s(k_{\ast})\approx 10^{-2}$, and $\beta_s(k_{\ast})\approx 10^{-2}$.

 FIG. 1 presents the angular power spectra, $C_l$, for the CMB temperature anisotropy. In this Figure, the solid black curve represents the $C_l$ for the $\Lambda$CDM model with the best fit parameters obtained using the Planck TT + lowP data. The red dashed curve represents the $C_l$ for the Starobinsky potential. To obtain this curve, we have changed the CAMB to contain the scalar and tensor power spectrum related to the Starobinsky model which was  calculated in previous section using the best fit value of $N_{\ast}=63$. We have also plotted the CMB Planck 2015 data for comparison.
  As shown in FIG. 1, the suppression of $C_l$ predicted by the Starobinsky model for $l<50$ is the most important factor in reducing the likelihood compared to the $\Lambda$CDM model.

\section{Conclusion }

In inflationary models, it is important to know how the model predictions can be compared with observational data. A variety of methods have been proposed in the literature for such comparisons. In this work, we presented a new method for simple models of inflation. In this method, we first expanded the logarithm of scalar and tensor power spectra around the pivot scale to found the series terms for the single field inflation models as a function of $N_{\ast}$. The expansion terms are denoted by the spectral index, the running, the running of running, and so on. In order to find the series terms, we had to solve the equations of motion for the inflaton field, $\phi$, at the background to find $\phi$ as a function of the e-folding number, $N_{\ast}$, and to calculate the power spectra as a function of $N_{\ast}$. Using this approach, we were able to find the series terms in the final step. For the three types of single field inflation models with simple potentials studied here, we were able to evaluate the sums and find the spectral indexes and, thereby, the power spectra as a function of $k$ with one or two free parameters depending on the model. We compared such new power spectra with the standard power spectra of the $\Lambda$CDM model consisting of $n_{s}(k^{\ast})$, $\alpha_s(k^{\ast})$, $\beta_s(k^{\ast})$, and $r$ variables by using the modified version of CAMB and CosmoMC codes as well as the recent Planck TT + lowP and BKP data. The goodness of fit of the specific inflationary models with $N_{\ast}$ as the free parameter was found to be  comparable with that of the $\Lambda$CDM model for Planck TT + lowP and BKP data sets. In the case of Starobinsky model, the new parametrization yielded a  better fit.
Using this analysis, we were also able to constrain the e-folding number for any inflation model. For an instantaneous reheating stage, one expects the bound $44<N_{\ast}<59$. We compared the best fit values of $N_{\ast}$ calculated for different models using this bound. It was found that the models that predict the worst values of $\chi^{2}$ also predict the $N_{\ast}$ parameter outside of this bound.
 We expect that our method of analyzing the inflationary models can be extended to other models with more complex potentials.

\section*{\small Acknowledgment}

The author would like to thank H. Firouzjahi, M. S. Movahhed, A. Abolhasani, S. Baghram, M. H. Namjoo for various comments and useful discussions.


\end{document}